\documentclass[11pt]{article}
\usepackage{blois,epsfig}

\def\Journal#1#2#3#4{{#1} {\bf #2}, #3 (#4)}

\def\be{\begin{equation}}
\def\ee{\end{equation}}
\def\bea{\begin{eqnarray}}
\def\eea{\end{eqnarray}}

\def\laeq{\raise.2ex\hbox{$<$}\kern-.75em\lower.9ex\hbox{$\sim$}\,}
\def\gaeq{\raise.2ex\hbox{$>$}\kern-.75em\lower.9ex\hbox{$\sim$}\,}

\begin{document}
\vspace*{4cm}
\title{MODELING OF GAMMA-RAY MILLISECOND PULSAR\\ LIGHT CURVES REVEALED BY \textit{FERMI}-LAT}

\author{C. VENTER$^{1,2,3}$, A.~K. HARDING$^{1}$, and L. GUILLEMOT$^{4}$}

\address{$^1$Astrophysics Science Division, NASA Goddard Space Flight Center, Greenbelt, MD 20771, USA\\
$^2$Unit for Space Physics, North-West University, Potchefstroom Campus, Private Bag X6001, Potchefstroom 2520, South Africa\\
$^3$NASA Postdoctoral Program Fellow\\
$^4$CNRS/IN2P3, Universit\'{e} de Bordeaux, Centre d'\'{E}tudes Nucl\'{e}aires de Bordeaux Gradignan, UMR~5797, 33175 Gradignan, France}

\maketitle\abstracts{
\textit{Fermi} Large Area Telescope (LAT) has recently detected~8 gamma-ray millisecond pulsars (MSPs), providing an unprecedented opportunity to probe the magnetospheres of these low-spin-down pulsars. We performed 3D emission modeling, including various Special Relativistic effects, in the context of pair-starved polar cap (PSPC), slot gap (SG), and outer gap (OG) pulsar models. Most of the light curves are best fit by SG and OG models, surprisingly indicating the presence of narrow accelerating gaps limited by robust pair production. All model fits imply high-altitude emission, and we observe exclusive differentiation of the current gamma-ray MSP population into two sub-classes: light curve shapes and lags across wavebands impose either PSPC or SG / OG-type geometries.}
%In the first case, the radio pulse has a small lag with respect to the single gamma-ray pulse, while the (first) gamma-ray peak usually trails the radio by a large phase offset in the latter case.}

\section{Introduction}
\label{sec:intro}
Millisecond pulsars (MSPs) are rapidly-rotating neutron stars, characterized by relatively short spin periods and low surface magnetic fields. %Following its birth during a supernova explosion, a pulsar matures and spins down until cessation of radio emission occurs. 
It is thought that transfer of mass and angular momentum from a binary companion during an accretion phase, may revive their ``dead'' (radio-invisible) younger pulsar progenitors which have ``spun down'' previously.~\cite{Alpar82} MSPs are thus believed to be ``recycled'' pulsars,~\cite{Bhattacharya91} decending from low-mass X-ray binaries (LMXRBs).~\cite{Archibald09}

MSPs were predicted to be visible in gamma rays.~\cite{HUM05,Frackowiak05a,Zhang07,Venter09,Zajczyk08,Watters09,Venter09_MSPs} Observational confirmation was provided by the \textit{Fermi} Large Area Telescope (LAT), a high-energy (HE) gamma-ray telescope with a large field of view (2.4~sr), operating in the energy range $\sim20$~MeV to $>300$~GeV~\cite{Atwood09}, which recently detected~8 gamma-ray MSPs.~\cite{Abdo09_MSP}

%\textit{Fermi}-LAT has made many important pulsar discoveries, including detection of the radio-quiet gamma-ray pulsar inside the supernova remnant CTA~1,~\cite{Abdo08_CTA1} the~6 high-confidence \textit{EGRET} pulsars,~\cite{Thompson04} 16~radio-quiet pulsars using blind searches,~\cite{Abdo09_BS} and in particular, discovery of~8 gamma-ray MSPs.~\cite{Abdo09_MSP}

Mainly two different models
%, associated with different emission regions in the pulsar magnetosphere, 
have been employed to describe HE radiation from pulsars. Polar cap (PC) models~\cite{DH96} assume injection of primary electrons from the stellar surface. These primaries radiate HE curvature radiation (CR) or inverse Compton scattering (ICS) gamma rays which are converted into electron-positron pairs via magnetic pair production. A low-altitude pair formation front (PFF) is established close to the surface, screening the accelerating electric field.~\cite{HM98} By allowing variation of the CR PFF altitude across the PC, a slot gap (SG) is formed along the last open magnetic field lines. This gap extends from the neutron star surface to near the so-called light cylinder, and allows acceleration of primaries up to high altitudes.~\cite{Arons83,MH04_SG} The SG model is a possible physical realization of the two-pole caustic geometry.~\cite{Dyks03} %which was developed to explain pulsar HE light curves. 
Outer gap (OG) models~\cite{CHR86a,CHR86b,Romani96} assume that HE radiation is generated by cascades of electron-positron pairs produced via photon-photon pair creation. This emission occurs close to the last open field lines, above the ``null-charge surface'', and is thus confined to the outer magnetosphere. Recently, a 3D OG solution was found~\cite{Hirotani08} which extends toward the NS surface.
%, where a small acceleration field extracts ions from the stellar surface in a space-charge-limited-flow regime.
%A variation of the standard PC model applies to MSPs, which are believed to have mostly unscreened magnetospheres due to their relatively low B-fields which inhibit copious magnetic pair production. We will refer to it as a ``pair-starved polar cap'' (PSPC) model.~\cite{MH04_PS,HUM05,MH09_PS,Venter09_MSPs} 
For MSPs with mostly unscreened magnetospheres,  a ``pair-starved polar cap'' (PSPC) model~\cite{MH04_PS,Venter09_MSPs} applies in which charges are accelerated up to high altitudes over the full open-field-line region. %(An SG accelerator progressively increases in size with pulsar age, and relaxes to a PSPC structure in the limit of no electric field screening.)

We study the \textit{Fermi}-LAT MSP population~\cite{Abdo09_MSP} using 3D emission modeling, including Special Relativistic (SR) effects of aberration and time-of-flight delays, and rotational sweepback of B-field lines,~\cite{Morini83,Romani95,Dyks04_SR} in the context of geometric SG and OG models, as well as a CR PSPC model, and obtain fits for gamma-ray and radio light curves. Our calculations are complementary to another study~\cite{Watters09} that focuses on younger pulsars. %Section~\ref{sec:Model} describes the various models we have applied, and we present our results and conclusions in Sections~\ref{sec:Results} and~\ref{sec:Con}. 
More details are provided elsewhere.~\cite{Venter09_MSPs}

\begin{figure}
\begin{center}
\epsfig{figure=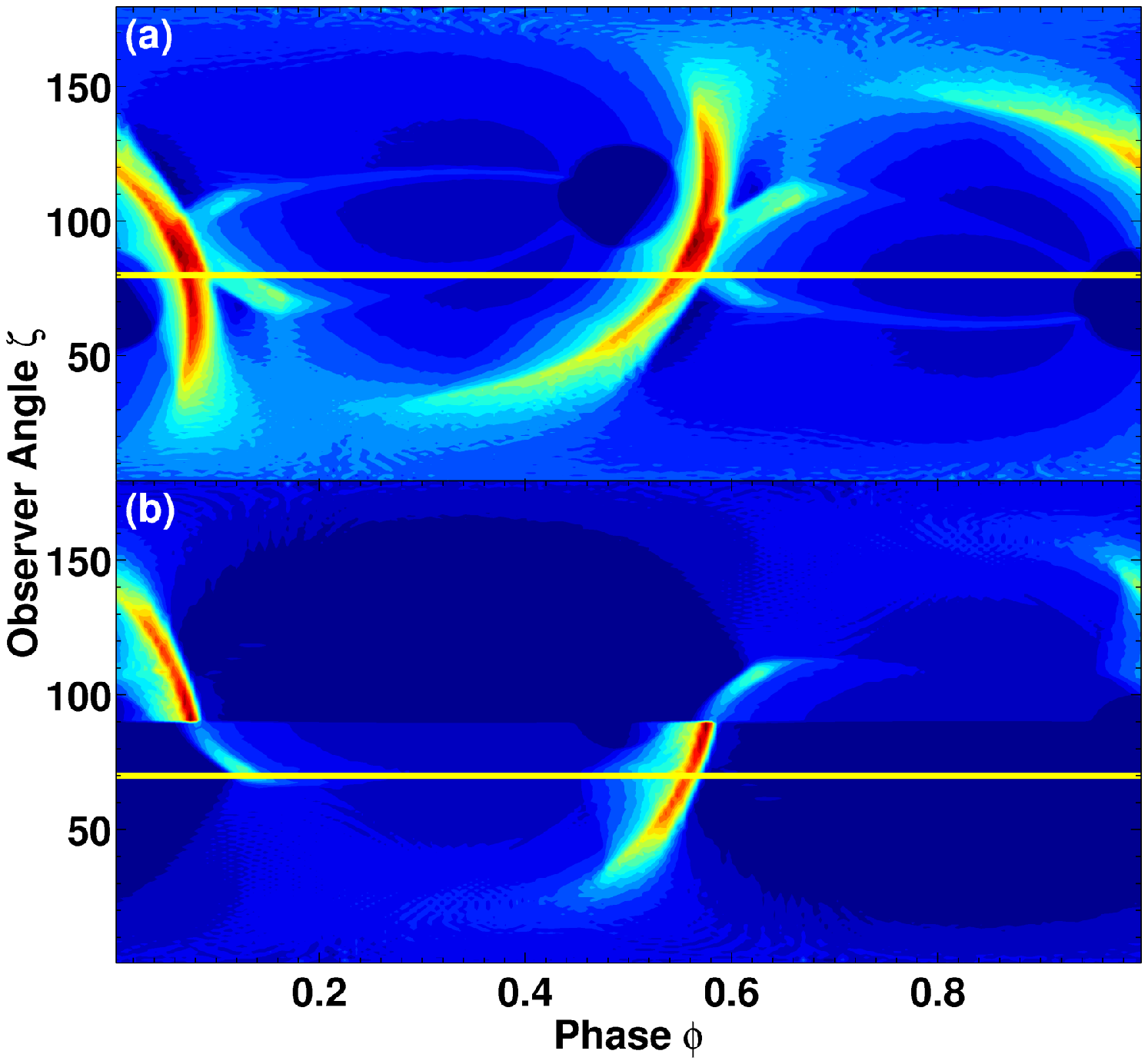,height=2.5in}\hskip0.4cm
\epsfig{figure=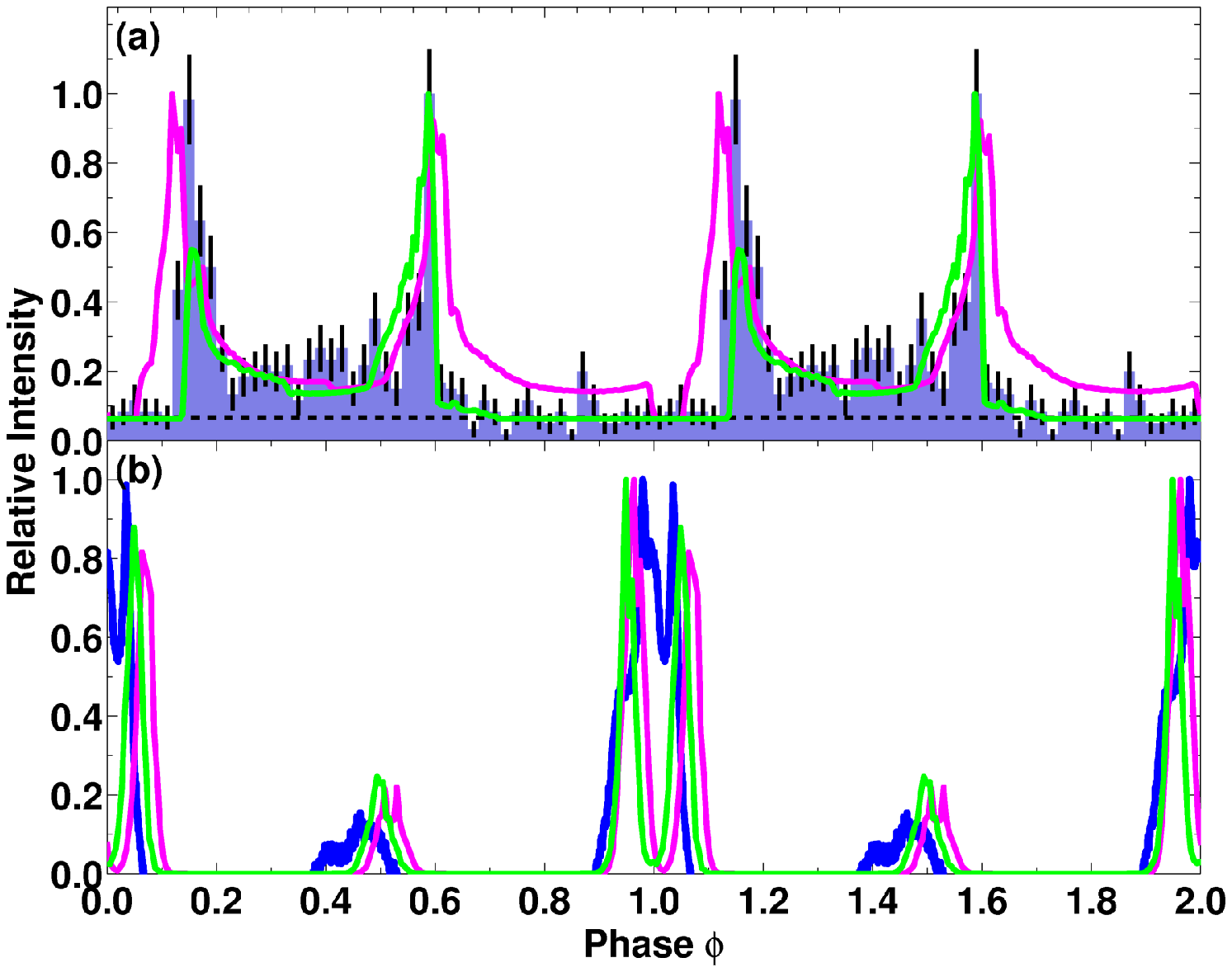,height=2.5in}
\caption{\textit{Left:} Phaseplots for PSR~J0030+0451. Panel~(a) is for an SG model with $(\alpha,\zeta)=(70^\circ,80^\circ)$, while panel~(b) is for an OG model with $(\alpha,\zeta)=(80^\circ,70^\circ)$. \textit{Right:} Observed and fitted light curves for PSR~J0030+0451. In panel~(a), we show the \textit{Fermi}-LAT data (histogram) with estimated background level (dashed line), and SG (magenta line) and OG (green line) fits. Panel~(b) shows the radio data (blue line), and the magenta and green lines correspond to the same $(\alpha,\zeta)$ combinations as those of the SG and OG fits in panel~(a).\label{fig1}}
\end{center}
\end{figure}

\section{Model Description}
\label{sec:Model}
%\subsection{B-field and SR Effects}
%\label{sec:B}
We use an implementation~\cite{Dyks04_B,Dyks04_Bovc} of the retarded vacuum dipolar B-field solution~\cite{Deutsch55} of a rotator inclined by an angle $\alpha$ with respect to the rotation axis.
%following earlier work~\cite{Romani95,Higgins97,Arendt98,Cheng00}. 
The PC shape is distorted asymmetrically by rotational sweepback of field lines. We calculate the curvature radius of the B-field lines in the inertial observer frame, and assume SG and OG gaps to be confined between two field lines with footpoints close to the PC rim, and having constant emissivity over the emitting volume. In the PSPC case, we collect photons with energies above 100~MeV from the full open-field-line region.
%\subsection{Particle Transport and PSPC E-field}
%\label{sec:Transport}

For the PSPC regime, we only consider CR losses suffered by electron primaries moving along the field lines. Previous studies~\cite{HUM05,Frackowiak05a,Venter09,Zajczyk08,VdeJ05} have used lower-altitude solutions~\cite{HM98,MH97} for $E_{||}$, the PSPC E-field parallel to the B-field. As \textit{Fermi}-LAT results seem to indicate that the HE radiation is originating in the outer magnetosphere,~\cite{Abdo09_MSP} we now incorporate the solution~\cite{MH04_PS} for altitudes close to the light cylinder in the small-angle approximation %(small $\alpha$, $\xi$, and high altitude), 
by matching this solution to the two lower-altitude ones. 
%Our matching procedure also mitigates the problem of particle oscillations which occurs when the E-field reverses sign beyond some altitude. 
We find conservation of energy when solving the transport equation for relativistic electron primaries using the full E-field solution.
%\subsection{Generation of Phaseplots}
%\label{sec:Phaseplots}
%In the case of the PSPC model, we normalize the particle outflow along each B-line according to $d\dot{N}(\xi,\phi_{\rm pc}) = -\rho_e(\eta=1,\xi,\phi_{\rm pc})dS\beta_0c/e$, with $d\dot{N}$ the number of particles leaving a surface patch $dS$ per unit time with initial speed $\beta_0c$ and $\rho_e$ is the charge density given by Eq.~(12) of \cite{HM98}. The latter is equal to the GR equivalent of the Goldreich-Julian charge density at the NS surface. We assume that $d\dot{N}$ has no radial dependence. We next distributed primary electrons uniformly across the PC, with electron positions coinciding with B-line footpoints on the stellar surface, and followed the motion of electron primaries along these lines (Section~\ref{sec:Transport}), collecting and binning HE CR. For all the other geometric models, we assume constant emissivity per unit length. We lastly accumulate the number of photons radiated per unit time in $(\zeta,\phi)$-bins, and divide by the solid angle subtended by each phaseplot bin to make up the final phaseplot.
%\subsection{Radio Beam Model}

We use an empirical radio cone model based on the characterization~\cite{Rankin93} of the radio emission as magnetic axis-centered core and conal beams. We adopt a description~\cite{Gonthier04} based on fits~\cite{ACC02} of average-pulse profiles of a small collection of pulsars at 400~MHz. The conal emission occurs at altitudes of $10\% - 20\%$ of the light cylinder radius. We lastly use relative units for the cone beam luminosity. %More details are described elsewhere.~\cite{Venter09_MSPs}

\section{Results}
\label{sec:Results}
%Both the PC area and the radio beam are significantly larger for MSPs than for younger pulsars. The notch, a feature of the retarded magnetic field solution \cite{Dyks04_B}, is generally also clearly distinguishable. 
%We assume that emission is generated tangent to the B-field lines (and symmetric for both magnetic poles). 
SR effects of aberration and time-of-flight delays cause phase shifts that nearly cancel those due to the curvature of the B-field on trailing field lines, leading to accumulation of emission (initially emitted tangentially to the B-field lines) around narrow phase bands $\phi$. This results in caustic structures (on phaseplots of $\zeta$ vs.\ $\phi$, with $\zeta$ the observer angle) around $\phi\sim0.0-0.1$ and $\phi\sim0.4-0.6$ in phase for SG and OG models. No emission originates below the null charge surface %(at $\zeta=90^\circ$ in phaseplots) 
in the OG model, so that an observer can only see emission from one magnetic pole, in contrast to the SG models.
%The constant-emissivity PC vs.\ PSPC models differ for two reasons: (i) The PSPC E-field decreases with altitude before reaching a constant value, leading to much smaller emission regions, and also relatively less high-altitude emission. (ii) The PSPC E-field depends on spatial parameters, influencing the beam shape.

We studied a large number of light curves for each of the different pulsar models. Both OG and SG models have a preponderance of double-peaked light curves at similar phases. OG models do not exist at all angle combinations, while SG models do (due to emission occuring below the null charge surface). One may find sharp, solitary peaks for some regions in $(\alpha,\zeta)$-space in OG models, while the corresponding SG-peaks usually have additional low-level features. 
%, and for nearly the full range of inclination and observer angles ($\alpha = \zeta = 5^\circ-90^\circ$, in $5^\circ$ intervals); also for $P=2$, 3, and~5~ms, and for different gap widths. 
The PSPC model have mostly single-peaked gamma-ray profiles roughly in phase with the (single) radio peaks, but one may see radiation from both poles for large~$\alpha$, leading to double-peaked profiles separated by $\sim0.5$ in phase.
%, and which become larger when $P$ decreases (especially for the radio profiles). 
The radio peak multiplicity depends on the observer's geometry. In significantly off-beam geometries, only the relatively larger gamma-ray cone is seen, explaining the phenomenon of ``radio-quiet'' pulsars. 

We chose best-fit light curves from the various models to match the MSP gamma-ray and radio data by eye. %However, statistical uncertainties may complicate unique matching of predicted and observed profiles. In addition, the model light curves usually do not radically change for a $\sim5^\circ$-change in $\alpha$ or $\zeta$, making our obtained fits somewhat subjective. 
As an example, the left panel of Figure~\ref{fig1} shows phaseplots %(number of photons per solid angle vs.\ observer angle $\zeta$ vs.\ phase $\phi$) 
associated with the best light curve fits (right panel) %to the observed gamma-ray and radio data 
obtained for PSR~J0030+0451. %The upper left subpanel indicates an SG model, and the lower left one an OG model. 
%We did not find any satisfactory fits from the geometric TPC and PC models. 
We inferred values for $\alpha$ and $\zeta$ for each MSP (Table~\ref{tab2}), and compared them with (somewhat uncertain) values obtained from radio polarimetric measurements (references for radio data elsewhere~\cite{Venter09_MSPs}). %The latter values are typically very uncertain for MSPs due to the flatness of the position angle curve, and scatter of data.

%Three MSPs (PSR~J0030+0451, PSR~J0218+4232, and PSR~J1614$-$2230) have double-peaked light curves, indicating the presence of screening electron-positron pairs (which are are necessary to form the SG or OG emitting structure). In six cases, the gamma-ray light curve lags the radio. Two MSPs, PSR~J0030+0451 and PSR~J1614$-$2230, have a relative phase lag $\delta\sim0.2$ (distinct from the function $\delta(\eta)$ used earlier), and four, PSR~J0218+4232, PSR~J0437$-$4715, PSR~J1613$-$0200, and PSR~J0751+1807, have $\delta\sim0.45$. These MSPs are well fit by SG and OG models. The remaining two MSPs (PSR~J1744$-$1134 and PSR~J2124$-$3358) have $\delta\sim0.85$, which means that the radio lags the gamma-ray curves by 0.15 in phase. These two cases are exclusively fit by the PSPC model, where the gamma and radio emission come from the same magnetic pole, and the gamma and radio emission originate well above the stellar surface. For PSR~J1614$-$2230, the radio profile was measured at 1.5~GHz, and for PSR~J0437-4715, at 3~GHz (although for the modeling we only use frequencies 1.4~GHz and 3~GHz). All other radio profiles were observed at 1.4~GHz \cite{Abdo09_MSP}. 

\begin{table}[t]
\caption{Model fits for inclination and observer angles $\alpha$, $\zeta$.\label{tab2}}
\vspace{0.4cm}
\begin{center}
\begin{tabular}{|c|c|c|c|c|c|c|c|c|}
\hline
& & & & & & & & \\
Name & $\alpha_{\rm SG}$ & $\zeta_{\rm SG}$ & $\alpha_{\rm OG}$ & $\zeta_{\rm OG}$ & $\alpha_{\rm PSPC}$ & $\zeta_{\rm PSPC}$ & $\alpha_{\rm radio}$ & $\zeta_{\rm radio}$ \\
     & ($^\circ$) & ($^\circ$) & ($^\circ$) & ($^\circ$) & ($^\circ$) & ($^\circ$) & ($^\circ$) & ($^\circ$)\\
\hline
J0030+0451   & 70      & 80      & 80      & 70      &   &   & $\sim62$      & $\sim72$\\          %& \cite{Lommen00}      \\
J0218+4232   & 60      & 60      & 50      & 70      &   &   & $\sim8$       & $\sim90$\\          %& \cite{Stairs99}      \\
J0437$-$4715 & 30      & 60      & 30      & 60      &   &   & $20-35$       & $16-20$\\           %& \cite{MJ95,Gil97}    \\
J0613$-$0200 & 30      & 60      & 30      & 60      &   &   & small $(\zeta-\alpha)$ &       \\      %& \cite{Xilouris98}    \\
J0751+1807   & 50      & 50      & 50      & 50      &   &   &         &            \\ %&     	 	    \\
J1614$-$2230 & 40      & 80      & 40      & 80      &   &   &         &            \\ %&     		    \\
J1744$-$1134 &   &   &   &   & 50      & 80      &         &            \\ %&     	    \\
J2124$-$3358 &   &   &   &   & 40      & 80      & $20-60$ (48)  & $27-80$ (67)     \\ %& \cite{Manchester04} \\
\hline
\end{tabular}
\end{center}
\end{table}

\section{Discussion and Conclusions}
\label{sec:Con}
We have compared 3D model predictions of gamma-ray and radio radiation with MSP gamma-ray data from \textit{Fermi}-LAT in the framework of geometric SG and OG pulsar models, and also for the full-radiation PSPC model. Surprisingly, some MSPs have double-peaked light curves well fit by SG / OG models, indicating strong screening of $E_{||}$ by pair production. The larger radio beam widths of MSPs compared to those of canonical pulsars furthermore implies relatively few radio-quiet MSPs.
%New ways of creating pairs in low-$\dot{E}_{\rm rot}$-pulsars will have to be found to explain this phenomenon. 
We found exclusive differentiation between the SG / OG models on the one hand, and the PSPC model on the other hand. %, by producing both the gamma-ray and radio curves within the same model.
%Six MSPs have gamma-ray light curves which lag the radio and are explained using SG or OG fits, but not PSPC fits. For the remaining two MSPs, the radio curves slightly lag the gamma-ray curves, and these are fit by the PSPC model (and not by the SG / OG models). 
Emission in \textit{all} models considered comes from the outer magnetosphere. Our fits of $\alpha$ and $\zeta$ are in reasonable agreement with values inferred from MSP radio polarization measurements.
For PSR~J0437-4715 and PSR~J0613-0200, the SG model predicts a small precursor to the main gamma-ray peak, but not the OG model. This may become a future model discriminator.
Future phase-resolved spectroscopy made possible by the quality of \textit{Fermi}-LAT data should challenge pulsar models.
% to reproduce such unprecedented detail.

\section*{Acknowledgments}
CV is supported by the NPP at the Goddard Space Flight Center, administered by ORAU through a contract with NASA, and also by the South African National Research Foundation. AKH acknowledges support from the NASA Astrophysics Theory Program. %We thank Alex Muslimov and Jarek Dyks for useful discussions.

\end{document}